\documentclass[useAMS,usenatbib]{mn2e}
\usepackage{amsmath,fleqn,graphicx,amssymb,xcolor}
\usepackage{multirow}
\renewcommand{\[}{\begin{equation}}
\renewcommand{\]}{\end{equation}}
\def\p{\partial}\def\i{{\rm i}}

\let\boldgrk=\gkvecten
\let\boldgrksc=\gkvecseven

\def\gkthing#1{{\mathchoice%
	{\hbox{{\boldgrk\char#1}}}
	{\hbox{{\boldgrk\char#1}}}
	{\hbox{{\boldgrksc\char#1}}}
	{\hbox{{\boldgrksc\char#1}}}}}

\def\vepsilon{\gkthing{15}}

\def\vtheta{\gkthing{18}}

{\newif\ifnotend
\notendtrue
\def\veclist{ABCDEFGHIJKLMNOPQRSTUVWXYZabcdefghijklmnopqrstuvwxyz.}
\def\top#1#2.{#1}
\def\tail#1#2.{#2.}
\loop\expandafter\xdef\csname v\expandafter\top\veclist\endcsname%
{{\noexpand\bf\expandafter\top\veclist}}
\edef\veclist{\expandafter\tail\veclist}
\if\veclist.\notendfalse\fi\ifnotend\repeat}
{\newif\ifnotend
\notendtrue
\def\callist{ABCDEFGHIJKLMNOPQRSTUVWXYZ.}
\def\top#1#2.{#1}
\def\tail#1#2.{#2.}
\loop\expandafter\xdef\csname c\expandafter\top\callist\endcsname%
{{\noexpand\cal\expandafter\top\callist}}
\edef\callist{\expandafter\tail\callist}
\if\callist.\notendfalse\fi\ifnotend\repeat}
\def\rhop{\rho^{(\alpha)}}
\def\Phip{\Phi^{(\alpha)}}\def\Phipps{\Phi^{(\alpha')*}}

\def\rhopp{\rho^{(\alpha')}}\def\Phips{\Phi^{(\alpha)*}}
\def\ex#1{\left<#1\right>}
\def\bra#1{\langle#1|}
\def\ket#1{|#1\rangle}
\def\la{\langle}
\def\ra{\rangle}
\def\cEx{\cE_\xi }
\def\real{\Re\hbox{{\rm e}}}                   
\def\d{{\rm d}}

\def\bolOm{\mbox{\boldmath$\Omega$}}
\def\vOmega{\bolOm}

\def\e{\mathrm{e}}

\def\fracj#1#2{{\textstyle{#1\over#2}}}

\def\eqrf#1{(\ref{#1})}

\title[Modes of a stellar system I: ergodic systems]
{Modes of a stellar system I: ergodic systems}

\author[Jun Yan Lau \& James Binney]{
  Jun Yan Lau$^{1,2}$\thanks{E-mail: jun.lau.20@ucl.ac.uk} \& James Binney$^2$\thanks{E-mail:
  binney@physics.ox.ac.uk}\\   
  $^1$UCL Mullard Space Sciences Laboratory, Holmbury St Mary, Surrey RH5
  6NT\\
  $^2$Rudolf Peierls Centre for Theoretical Physics, Clarendon Laboratory,
  Parks Road, Oxford, OX1 3PU, UK
}

\begin{document}
\maketitle

\begin{abstract}
The excursions of star clusters and galaxies around statistical equilibria
are studied. For a stable ergodic model Antonov's Hermitian operator on
six-dimensional phase space has the normal modes as its eigenfunctions. The
excitation energy of the system is just the sum of the (positive) energies
associated with each normal mode. Formulae are given for the DFs of modes,
which are of the type first described by van Kampen rather than Landau, and
Landau `modes' can be expressed as sums of van Kampen modes.  Each van Kampen
mode
comprises the response of non-resonant stars to driving by the gravitational
field of stars on a group of resonant tori, so its structure is
sensitive to the degree of self gravity. The emergence of global distortions
in N-body models when particles are started from an analytical equilibrium is
explained in terms of the interplay of normal modes.  The positivity of modal
energies opens the way to modelling the thermal properties of clusters in
close analogy with those of crystals.
\end{abstract}


\section{Introduction} \label{sec:intro}

Galaxies and star clusters are in approximate states of equilibrium and have
for decades been fitted to models in which the distribution function
$f(\vx,\vv)$ of their constituent particles (stars, dark-matter particles)
are steady-state solutions of the collisionless Boltzmann equation (CBE).
The advent of massive simulations of galaxy formation \citep{Laporte2019} and
detailed data from the Gaia mission \citep{GaiaDR2general} and large integral
field units such as MUSE \citep[e.g.][]{MUSE2021} have stimulated interest in
non-equilibrium features of galaxies, especially the Milky Way
\citep{AntojaSpiral}. 

For almost a century observations of galaxies and star clusters have been
interpreted in terms of `mean-field' models, that is to say models in which
fluctuations have been averaged away. In the case of globular clusters the
community has been aware since at least the pioneering work of
\cite{Henon1961}
that fluctuations drive secular evolution of the system towards higher central
concentration and lower mass (core collapse and evaporation) but
observations have nonetheless been fitted to mean-field models on the grounds
that clusters evolve through a series of mean-field models.

Fluctuations in the surface brightnesses of early-type galaxies form the
basis for a standard technique for estimating their distances
\citep{TonrySchneider}, but the fluctuations are computed by imposing shot
noise on equilibrium models rather than using a dynamical theory of
fluctuations.

Perhaps the most exciting single discovery made in the Gaia DR2 data is the
phase spiral that \cite{AntojaSpiral} uncovered in the distribution of stars in
the $(z,v_z)$ plane. The spiral is surely a symptom of a macroscopic
oscillation of the disc that has a significant component in the $z$
direction. If we had a credible dynamical model of this oscillation, we would
be able to extract from the Gaia data information about the structure of the
disc and the agent [likely the Sagittarius dwarf
galaxy \citep{BinneySchoenrich2018,Laporte2019,JBHThor2020}] that excited it.
Unfortunately, the disc's self-gravity certainly plays an important role in
the oscillation, and there's little prospect of adequately modelling the
disc's oscillation until  we have a better understanding of the global
oscillations of stellar systems. This is the first in a series of papers
that lay the foundations for such understanding by setting up an adequate
theory of the normal modes of stellar systems.

Normal modes (in quantum mechanics `stationary states') owe their usefulness
to three key properties: (i) they are complete in the sense that any initial
condition can be expressed as a linear combination of normal modes; (ii) they
have the trivial time dependence $\e^{-\i\omega t}$; (iii) they are
mutually orthogonal, with the consequence that the energy of the whole system
is simply the sum of the energies invested in each normal mode. 

Modes of stellar systems have received significant attention since the work
of \cite{To64}, \cite{LinShu}, and \cite{Kalnajs1965}. That work
was motivated by the desire to understand spiral structure so focused on
razor-thin, rotating stellar discs. Two decades later the focus switched to
hot, spherical systems from a desire to understand how and when radial bias
in the velocity dispersion caused systems to lose spherical symmetry
\citep{PalmerPapaloizou1987,Saha1991,Weinberg1991}.  The standard reference for this work is
the two volumes of \cite{FridmanPolyachenkoI}, and a glance at the contents
pages make clear that interest focused exclusively on the search for unstable
normal modes. We show below that these modes are qualitatively different from
the modes required to investigate, as we do, the excursions that stable
systems make around equilibrium. 

Fluctuations may be externally or internally driven. The Antoja spiral in our
Galaxy and shells around early-type systems \citep{MalinCarter1980} are
surely externally driven. The secular evolution of globular clusters is
largely driven by fluctuations that are internally driven by Poisson noise
[although fluctuations driven externally by tidal fields are also
significant \citep{LeeOstriker1987}]. Even after more than a half century of work, there is no
consensus as to whether observed spiral structure is sometimes internally
driven (Sellwood \& Masters 2022), although some `grand-design' spiral structure
(e.g., that of M51) is certainly externally driven. Whatever the  driving
mechanism, the natural way to model fluctuations is as solutions to the
linearised Boltzmann equation (CBE) coupled to the already linear Poisson
equation. 

This time-translation invariant pair of linear equations may be
expected to have a complete set of solutions with time dependence
$\e^{-\i\omega t}$ (with potentially complex $\omega$). In this paper we
derive these solutions for the important special case that the unperturbed
system is ergodic -- that is has a distribution function (DF) of the form
$f_0(H)$, where 
\[
H(\vx,\vv)=\fracj12 v^2+\Phi(\vx)
\]
is the Hamiltonian of a single particle moving in the gravitational potential
$\Phi$. The second paper in the series generalises many results to the case
of a DF of the form $f_0(\vJ)$, where $\vJ$ is the vector of the action
integrals of stars moving in the unperturbed potential.  In the third paper
in the series we develop an apparatus for decomposing an arbitrary initial
condition of a system with $f_0(\vJ)$ into its constituent normal modes. The
present
paper relies heavily on an Hermitian operator that \cite{Antonov1961}
introduced. This operator does not generalise straightforwardly from ergodic
systems to more general ones, so Paper II obtains a restricted range of
results with a simpler but less powerful technique.  The
fourth paper in the series generalises Antonov's operator to DFs of the form
$f(\vJ)$.

The plan of this paper is as follows. Section \ref{sec:math} introduces basic
concepts and establishes notation.  Section \ref{sec:Ksec} introduces the
Hermitian operator $K$ on phase space whose eigenfunctions are the required
normal modes of the cluster.  If the system is stable, all its modes are van
Kampen modes; they have real frequencies drawn form a continuous spectrum. If
the system is unstable the spectrum contains isolated pure imaginary
frequencies. We show that the energies of modes are additive, and give a very
simple expression for a mode's energy in terms of its DF. This expression
implies that the energy of van Kampen modes is positive and that of modes
with imaginary frequencies vanishes. We show also that $K$ gives
rise to a slightly different conserved quantity that provides a means to
establish stability. In Section \ref{sec:Kcomm} we show that 
$K$ commutes with the angular-momentum operator $L_z$ before in 
Section \ref{sec:DF} obtaining an expression for
the DF of a van Kampen mode.  This contains a free function and parameters
that can be computed from the free function by matrix algebra.  In Section
\ref{sec:dress} we investigate the way in which the structures of a van Kampen
mode depends on the extent to which a system is self-gravitating.  In Section
\ref{sec:innerProd} we re-express a van Kampen mode's energy in terms of the
free function and the potential that the mode generates, and in Section
\ref{sec:Nbody} we discuss the emergence of system-scale fluctuations in N-body
simulations. In Section \ref{sec:complete} we argue that van Kampen modes
rest on conceptual foundations which are as solid as those that are generally
accepted as secure in other branches of physics.  In Section
\ref{sec:discuss} we stress the importance of the concept of particle
dressing in stellar dynamics as in other branches of physics, and discuss the
role that van Kampen modes play in dressing. Section \ref{sec:vK_Landau}
discusses the relationship between van Kampen and Landau modes, while Section
\ref{sec:thermo} considers the prospect for using van Kampen modes to extend
conventional statistical mechanics to stellar systems, and for understanding
the role of thermal fluctuations within them. Section \ref{sec:conclude} sums
up.

\section{Mathematical background}\label{sec:math}

Here we introduce essential mathematical tools and establish our notation.
We focus on stable ergodic clusters, that is systems with
unperturbed DFs $f_0(H)$ where
\[\label{eq:defHp}
H(\vx,\vv),\equiv\fracj12
v^{2}+\Phi(\vx)
\]
is the Hamiltonian of a single particle in the gravitational potential
$\Phi(\vx)$. A necessary and sufficient condition for such a system to be
stable is that the derivative $f_0'<0$ at all energies \citep{Antonov1961}. 

\subsection{Variable degree of self gravity}
In the following it proves helpful to be able to consider self-gravity to be
a variable $\xi $ that runs from zero (stars move in the fixed potential of a
specified density distribution) to unity (stars experience only their
gravitational attraction to the other stars). It is straightforward to set up
a numerical experiment for any given value of $\xi $ by sampling an analytic
density distribution in the usual way and taking the force on each star to be
$\xi $ times the force returned by an N-body solver plus $(1-\xi )$ times the force
provided by the analytic density.

\subsection{Angle-action variables} The role that Cartesian variables play for
homogeneous systems is played for spheroidal systems by angle-action variables
$(\vtheta,\vJ)$. The actions $J_i$ are constants of motion while their
conjugate variables, the angles $\theta_i$, increase linearly in time, so
$\vtheta(t)=\vtheta(0)+\vOmega t$. The particles' Hamiltonian $H(\vx,\vv)$ is a
function $H(\vJ)$ of the actions only and the frequencies $\Omega_i$ that
control the rates of increase of the angles are given by $\vOmega=\p
H/\p\vJ$. Angle-action variables are canonical, so the volume element of
phase space $\d^6\vw=\d^3\vx\d^3\vv=\d^3\vtheta\d^3\vJ$ and Poisson brackets can be computed
as
\[
[f,g]=\sum_i\bigg(
{\p f\over\p\theta_i}{\p g\over\p J_i}-{\p f\over\p
J_i}{\p g\over\p\theta_i}\bigg).
\]
Functions on phase space can be expressed as Fourier series:
\[\label{eq:defsFT}
h(\vw)=\sum_\vn h_\vn(\vJ)\e^{\i\vn\cdot\vtheta}\ ;\ 
h_\vn(\vJ)=\int{\d^3\vtheta\over(2\pi)^3}\e^{-\i\vn\cdot\vtheta}h(\vw).
\]
Note that for real $h$, $h_{-\vn}=h_\vn^*$.

\subsection{Potential-density pairs} Unfortunately, while the potential
$\Phi(\vx)$ is a function of only $\vx$, it becomes a function of both
$\vtheta$ and $\vJ$. So while angle-action variables make dynamics trivial,
they seriously complicate the solution of Poisson's equation.  Following
\cite{Kalnajs1976} this difficulty is finessed by introducing a basis of
{biorthogonal potential-density} pairs.  That is, a set of pairs
$(\rhop,\Phip)$ such that
\[\label{eq:Poisson}
4\pi G\rhop=\nabla^2\Phip\quad\hbox{and}\quad
\int\d^3\vx\,\Phips\rhopp =-\cE\delta_{\alpha\alpha'},
\]
 where $\cE$ is an arbitrary constant with the dimensions of energy.
Given a density
distribution $\rho(\vx)$, we expand it in the basis
\[\label{eq:defsAp}
\rho(\vx)=\sum_\alpha A_\alpha\rhop(\vx)\quad\Leftrightarrow\quad
\Phi(\vx)=\xi\sum_\alpha A_\alpha\Phip(\vx),
\]
where 
\[\label{eq:Aalpha}
A_\alpha=-{1 \over\cE}\int\d^6\vw\,\Phips(\vx)f(\vw).
\]
 If $\rho$ and $\Phi$ are time-dependent, the $A_\alpha$ become
time-dependent.  From equations \eqrf{eq:Poisson} and \eqrf{eq:defsAp} one
can obtain an expression for $\Phi$ in terms of $\rho$. Comparison of this
relation with Poisson's integral, yields
\[\label{eq:Gxmx}
{G\over|\vx'-\vx|}={1\over\cE}\sum_\alpha\Phip(\vx)\Phips(\vx').
\]

\section{Antonov's  operator K}\label{sec:Ksec}

We now derive for stable ergodic clusters the Hermitian operator
$K$ introduced by \cite{Antonov1961}.  The true normal modes of the system
are eigenfunctions of $K$ with non-negative eigenvalues that prove to be the
squares of the modes' (real) frequencies.
Following \cite{Antonov1961} we  split the perturbed DF $f_1$ into parts that are even and odd in $\vv$:
\[\label{eq:fSplit}
f_1(\vx,\vv)=f_+(\vx,\vv)+f_-(\vx,\vv)
\]
where
\[
f_\pm(\vx,\vv)\equiv\fracj12[f_1(\vx,\vv)\pm f_1(\vx,-\vv)].
\]
In the absence of a perturbation,  $f_-$ vanishes, so this part of the DF
isolates the effect of the perturbation. On the other hand the perturbation
changes the potential only through $f_+$.\footnote{ In Dirac's seminal
textbook, he argues that the second-order Klein-Gordon equation cannot stand
in for the Schr\"odinger equation, which is first-order in time. So he
factorises the Klein-Gordon operator into two first-order
operators, by splitting the wavefunction into four parts. Antonov
proceeded in the opposite direction: by splitting the DF he derived two
first-order operators and then combined them to obtain a second-order
equation.}

We define   an inner product on the space of DFs by
\[\label{eq:defIProd}
\la g\ket f\equiv-\int\d^6\vw\,{g^*f\over f'_0(H)},
\]
 where $H(\vJ)$ is the unperturbed Hamiltonian and the leading minus reflects the fact that $f'_0<0$.  Since $\d^6\vw\,f$
has dimensions of mass and $f/f'_0$ has dimensions of $v^2$, $\la g\ket f$
has the dimensions of $M v^2$, i.e., energy. When we Fourier expand the DFs
we find
\[
\la g\ket f=-(2\pi)^3\int{\d^3\vJ\over f_0'}\sum_\vk  g^*_{\vk} f_\vk.
\]
Notice that
\[\label{eq:splitIP}
\la f_1\ket{f_1}=\la f_-\ket{f_-}+\la f_+\ket{f_+}.
\]

$H(\vJ)$ is even in $\vv$ and the
Poisson bracket operator is odd in $\vv$, so our division \eqrf{eq:fSplit} of
$f$ splits the linearised CBE,
$\p_tf_1+[f_1,H]+[f_0,\Phi_1]=0$, into two equations
\[\label{eq:splitfdot}
{\p f_+\over\p t}=-[f_-,H]\quad;\quad
{\p f_-\over\p t}=-[f_+,H]+[\Phi_1,f_0].
\]
Now $[\Phi_1,f_0]=f_0'(H)[\Phi_1,H]$, so
\begin{align}
{\p\over\p t}[\Phi_1,f_0]&(\vw)=f_0'(H)\left[{\p\Phi_1\over\p t},H\right]\cr
&=-f_0'(H)\left[\int\d^6\vw'\,{\xi G\over|\vx'-\vx|}{\p f_+(\vw')\over\p
t},H(\vw)\right].
\end{align}
 We differentiate the second of equations \eqrf{eq:splitfdot} wrt $t$ and use the first
equation to eliminate $\p_tf_+$ from the rhs to obtain
\begin{align}
&{\p^2f_-\over\p
t^2}=[[f_-,H],H]\cr
&\quad+f_0'(H)\!
\left[\int\d^6\vw'\,{\xi G\over|\vx'-\vx|}[f_-(\vw'),H(\vw')],H(\vw)\right].
\end{align}
This equation is of the form
\[\label{eq:defKb}
{\p^2 f_-\over\p t^2}=-Kf_-,
\]
where the operator
\begin{align}\label{eq:defK}
K&\equiv-[[f_-,H],H]\cr
&-f_0'(H)\!
\left[\int\d^6\vw'\,{\xi G\over|\vx'-\vx|}[f_-(\vw'),H(\vw')],H(\vw)\right].
\end{align}

In terms of angle-action coordinates $(\vtheta,\vJ)$, 
\[\label{eq:PoissonHzero}
[f(\vtheta,\vJ),H(\vJ)]=\vOmega\cdot{\p f\over\p\vtheta},
\]
so {$K$ can be written
\begin{align}\label{eq:ddotfeK}
K&=-\left(\vOmega\cdot{\p\over\p\vtheta}\right)^2f_-\cr
&-f_0'(H)\,\vOmega\cdot{\p\over\p\vtheta}\int\d^6\vw'\,{\xi G\over|\vx'-\vx|}\vOmega'\cdot{\p
f_-(\vw')\over\p\vtheta'}.
\end{align}
Inserting equation \eqrf{eq:Gxmx}  into \eqrf{eq:ddotfeK} yields
\begin{align}\label{eq:ddotfeKb}
K&=-\left(\vOmega\cdot{\p\over\p\vtheta}\right)^2f_-
-f_0'(H){\xi\over\cE}\vOmega\cdot{\p\over\p\vtheta}\cr
&\times\int\d^6\vw'
\sum_\alpha\Phip(\vx)\Phips(\vx')\vOmega'\cdot{\p
f_-(\vw')\over\p\vtheta'}.
\end{align}

At this point it's convenient to define
\[\label{eq:defjalpha}
j_\alpha[f_1](t)\equiv-{\i\over\cE}\int\d^6\vw\,\Phips(\vx)\vOmega\cdot{\p
f_-(\vw)\over\p\vtheta},
\]
because it allows us to write equation \eqrf{eq:ddotfeKb} in the form
\[\label{eq:fddwithj}
K=-\left(\vOmega\cdot{\p\over\p\vtheta}\right)^2f_-
-\i\xi f_0'(H)\,\vOmega\cdot{\p\over\p\vtheta}
\sum_\alpha\Phip(\vx)j_\alpha[f_-].
\]
 $j_\alpha$ is a functional of $f_1$ rather than just $f_-$ because it can
also be computed from $f_+$: eliminating $f_-$ using the first of equations
\eqrf{eq:splitfdot} and equation \eqrf{eq:PoissonHzero} we obtain 
\[\label{eq:jf1t}
j_\alpha[f_1](t)=\i{1\over\cE}\int\d^6\vw\,\Phips(\vx){\p f_+\over\p t}.
\]
Since by equation \eqrf{eq:Aalpha} the coefficient $A_\alpha$ of the
potential/density expansion is a linear functional of $f_1$, and
$A[f_-]=0$, the derivative of $f_+$ in equation \eqrf{eq:jf1t} can be replaced by
a derivative of $A_\alpha$ to yield
\[\label{eq:jA}
j_\alpha[f_1](t)=-\i{\p A_\alpha\over\p t}.
\]

\subsection{Energy of a disturbance}\label{sec:E}

When we use equation \eqrf{eq:Gxmx} to eliminate $|\vx-\vx'|$ from equation
(5.130)
of \cite{GDII}, we find that the energy associated with a linearised disturbance is
\begin{align}\label{eq:BTE}
E[f_1]&=\fracj12\bigg\{\int{\d^6\vw\over|f_0'|}f_1^2
-{\xi\over\cE}\sum_\alpha\bigg|\int\d^6\vw\,\Phip(\vx)f_1(\vw)\bigg|^2\bigg\}\cr
&=\fracj12\Big\{\la f_-\ket{f_-}+\la
f_+\ket{f_+}-{\xi\cE}\sum_\alpha\big|A_\alpha[f_+]\big|^2\Big\}.
\end{align}
In the first line of this equation, the first and second terms on the right
quantify the potential and kinetic energies of the perturbation,
respectively. Hence the inner product $\la f_1\ket{f_1}$ gives twice a
perturbation's kinetic energy. In the last term, $A_\alpha$ is independent of
the degree of self-gravity $\xi $, so the final term is proportional
to $\xi$ as it should be.

In Appendix \ref{app:fKf} we show that $K$ is Hermitian and that
\[\label{eq:fmKfm}
\bra{f_-}K\ket{f_-}
=|\omega^2|\Big\{\la f_+\ket{f_+}-\xi\cE\sum_\alpha\big|A_\alpha[f_+]\big|^2
\Big\}.
\]
 When we use this equation to simplify equation \eqrf{eq:BTE}, we discover
the energy associated with an eigenfunction of $K$ identically vanishes in the
unstable case
$\omega^2<0$, and in the stable case is
\[\label{eq:EvK}
 E[f_1]=\la f_-\ket{f_-}=-(2\pi)^3\int{\d^3\vJ\over
 f_0'}\sum_\vk|f_{\vk-}|^2,
\]
where $f_{\vk-}$ denotes the $\vk$ component of $f_-$. Remarkably, the degree
of self-gravity $\xi$ does not appear in equation \eqrf{eq:EvK}.  Changes to
$\xi$ do affect $E$, however, by changing the $f_{\vk-}$.  The right
side of equation \eqrf{eq:EvK} is inherently positive and vanishes only if
$f_-$ vanishes. Moreover, by equation (\ref{eq:splitfdot}) $\omega
f_+=\vk\cdot\vOmega f_-$, so $f_+$ must also vanish if $f_-$ does. Hence {\it
the
energy of every mode of a stable ergodic system is positive.}

Since $K$ is Hermitian it has a complete set of orthogonal eigenfunctions.
Expressing an arbitrary disturbance as a linear combination of eigenfunctions
$f_-=\sum_\beta c_\beta f_-^{(\beta)}$ and inserting this expansion into
equation \eqrf{eq:EvK}, we conclude that the disturbance's energy is the sum of
the energies of its component eigenfunctions
\[\label{eq:sumE}
 E[f_1]=\sum_{\beta}|c_\beta|^2\la f_-^{(\beta)}\ket{f_-^{(\beta)}}.
\]

In view of these results it is natural to identify the {\it true modes} of an
ergodic system with the eigenfunctions of $K$. It follows that the
frequencies of true modes are either real or pure imaginary.  No true mode of
an ergodic system has negative energy; oscillatory modes have positive energy
and growing/decaying modes (if any) have zero energy.  Equation
\eqrf{eq:BTE}, from which we started, does not make evident the
non-negativity of energies.

\subsection{Antonov's conserved quantity}\label{secAntonov_E}

The rather involved argument just given starts from equation
\eqrf{eq:BTE} for the energy of a disturbance, which \cite{GDII} derive by
considering the work that must be done to initiate the disturbance. A much
simpler argument based on the Hermitian nature of $K$ yields the closely
related conserved quantity,
\[
\widetilde E=\fracj12\Big(\la\dot f_-\ket{\dot f_-}+\bra{f_-}K\ket{f_-}\Big).
\]
Indeed,
\[
2{\d\widetilde E\over\d t}
=\la\ddot f_-\ket{\dot f_-}+\la\dot f_-\ket{\ddot f_-}+\bra{\dot
f_-}K\ket{f_-}+\bra{f_-}K\ket{\dot f_-}=0.
\]
An eigenfunction of $K$ with eigenvalue $\omega^2$ has $\widetilde
E=\omega^2\la f_-\ket{f_-}$, so in this case conservation of $\widetilde E$
implies conservation of  $E=\widetilde E/\omega^2$. $\widetilde E$ does not
have the dimensions of energy, however.

Since $\la\dot f_-\ket{\dot f_-}>0$, instability, and thus systematic growth
of $f_-$, is excluded by conservation of $\widetilde E$ unless
$\bra{f_-}K\ket{f_-}<0$ for some function $f_-(\vx,\vv)$. In fact positivity
of $\bra{f_-}K\ket{f_-}$ for all functions in the natural space is a
necessary and sufficient condition for stability \citep{Laval1965,KulsrudMark}.

\section{DFs of van Kampen modes}\label{sec:vKDFsec}

If the
system is stable, the Hamiltonian's time-reversal invariance ensures that
$\omega$ is real because the existence of exponentially decaying solutions
would imply the existence of growing solutions. \cite{vanKampen1955} applied
these arguments to an electrostatic plasma and deduced some properties of the
normal modes of a plasma, which are known as van Kampen modes. The
corresponding modes of a stellar system have received little attention,
although \cite{Vandervoort2003} derived some of their properties. We now
examine the van Kampen modes of a stellar system in their role as
eigenfunctions of the operator $K$.

\subsection{Operators that commute with $K$}\label{sec:Kcomm}

Finding the eigenfunctions of $K$ is facilitated by identifying operators
that commute with $K$ and seeking eigenfunctions of $K$ that are also
eigenfunctions of
these operators.  $K$ is the operator associated with the time-translation
invariance of the underlying equilibrium. The Hamiltonian $H$ is invariant
under increments in the angle variables $\theta_i$ but in the presence of
self gravity ($\xi >0$), $K$ does not share the
invariance with respect to increments in $\theta_1$ and $\theta_2$ conjugate
to the radial action  $J_r$ and  the  modulus $L\equiv|\vL|$ of the angular
momentum vector $\vL$
because $\Phi_1(\vx)$ lacks this invariance.  Fortunately, $K$ is always invariant
under increments of the angle variable $\theta_3$ conjugate to $L_z$. To see
this it is best to return to  the definition of $K$ in equation
\eqrf{eq:defK}. Since $[H,L_i]=0$, we have
\begin{align}\label{eq:KcommL}
[K&f_-,L_z]=-[[[f_-,L_z],H],H]-f_0'(H)\cr
&\times
\bigg[\int\d^6\vw'\,\Big[{\xi G\over|\vx'-\vx|},L_z\Big]
[f_-(\vw'),H(\vw')],H(\vw)\bigg].
\end{align}
The operator $[.,L_z]$ rotates the orbital plane on which $\vx$ lies (by incrementing the argument
of the ascending node $\Omega$). The operator $[.,L_z+L_z']$, where $L_z'$ operates on $\vx'$, rotates $\vx$
and $\vx'$ through the same angle, so\footnote{The operator $[.,L]$ rotates the
orbit within the orbital plane while holding constant $\theta_1$, so in the
Keplerian case it rotates the ellipse rather than moving the star along its
ellipse. If $\vw$ and $\vw'$ lie on the same orbital plane, rotating both
orbits within the plane will leave $|\vx'-\vx|$ invariant, but if the points
lie on different planes,  $|\vx'-\vx|$ will not be invariant. Hence
$[|\vx'-\vx|,L]\ne0$.}
\[
[|\vx'-\vx|,L_z+L_z']=0.
\]
Hence taking advantage of the fact that for any $f,g,h$,
$\int\d^6\vw\,[f,g]h=\int\d^6\vw\,f[g,h]$, we have
\begin{align}
\int\d^6\vw'\,&\Big[{\xi G\over|\vx'-\vx|},L_z\Big][f_-(\vw'),H(\vw')]\cr
&=-\int\d^6\vw'\,\Big[{\xi G\over|\vx'-\vx|},L_z'\Big][f_-(\vw'),H(\vw')]\cr
&=\int\d^6\vw'{\xi G\over|\vx'-\vx|}\Big[[f_-(\vw'),H(\vw')],L_z'\Big]\cr
&=\int\d^6\vw'{\xi G\over|\vx'-\vx|}\Big[[f_-(\vw'),L_z'],H(\vw')\Big].
\end{align}
 When this result is
used in equation \eqrf{eq:KcommL}, we obtain
\[
\big[Kf_-,L_z]=K[f_-,L_z],
\]
so $K$ commutes with the operator $[.,L_z]$. From these commutations it
follows that the eigenfunctions of $A$ provide representations of the
group of translations around tori that is generated by $[.,L_z]$. This
compact Abelian group 
has only one-dimensional irreps, which can be reduced to multiplication
by $\e^{\i\alpha}$.  Hence the eigenfunctions of $A$ can be indexed by the integer
$n_3$ associated with $\theta_3$.

\subsection{Derivation of the DF}\label{sec:DF}

Bearing in mind that for an eigenfunction equation \eqrf{eq:jA}
yields $j_\alpha[f_1]=-\omega A_\alpha$, from equation \eqrf{eq:fddwithj} we have for an eigenfunction
\[
-\omega^2f_-=
\left(\vOmega\cdot{\p\over\p\vtheta}\right)^2\!\!f_-
-\i\xi\omega f_0'(H)\,\sum_\alpha A_\alpha\vOmega\cdot{\p\over\p\vtheta}
\Phip(\vx).
\]
 Now we apply the derivative $\vOmega\cdot\p/\p\vtheta$ to both sides and use
equations \eqrf{eq:splitfdot} and \eqrf{eq:PoissonHzero} to eliminate $f_-$ in favour of
$f_+$. Collecting terms with $f_+$ on the left we then have
\[
\left\{\omega^2+\left(\vOmega\cdot{\p\over\p\vtheta}\right)^2\right\}f_+=
\xi f_0'(H)\!\sum_\alpha A_\alpha\left(\vOmega\cdot{\p\over\p\vtheta}\right)^2\!\Phip(\vx).
\]
Fourier decomposed in angle variables this becomes
\[\label{eq:ModeEq}
\left\{\omega^2-(\vn\cdot\vOmega)^2\right\}f_{\vn+}=-
\xi f_0'(H)(\vn\cdot\vOmega)^2\sum_\alpha A_\alpha\Phip_\vn.
 \]
This equation is analogous to the standard equation for the
Laplace-transformed DF $\overline f\equiv\int_0^t\d t'\,f\e^{\i\omega t'}$
with $\Im(\omega)>0$,
\[\label{eq:oldftoPhi}
\i(\vn\cdot\vOmega-\omega)\overline f_\vn(\vJ,\omega)
=-f_0'(H)\vn\cdot\vOmega\sum_\alpha\overline A_\alpha(\omega)\Phip_\vn
+f_\vn(\vJ,0)
\]
in that it relates the
DF to the driving potential, but there are two significant  differences:

\begin{itemize}
\item[1] The rhs of equation \eqrf{eq:ModeEq} doesn't contain an  initial condition
analogous to  $\hat f_\vn(\vJ,0)$  on the rhs of \eqrf{eq:oldftoPhi}. It's not there because
we are seeking a normal mode rather than the solution to an initial-value
problem.

\item[2]
Equation  \eqrf{eq:ModeEq} starts with a factor $\omega^2-(\vn\cdot\vOmega)^2$
while equation \eqrf{eq:oldftoPhi} starts with  $\i(\vn\cdot\vOmega-\omega)$.
\end{itemize}

\noindent Before we divide equation \eqrf{eq:ModeEq} by $\omega^2-(\vn\cdot\vOmega)^2$, we must recognise that when it
vanishes, which for a range of frequencies it will over 2d {resonant surfaces} in action
space, $f_{\vn+}$ is unconstrained. This being so, after we've divided by
$\omega^2-(\vn\cdot\vOmega)^2$, we should add a function that's non-zero only
on resonant surfaces. Then we have
\begin{align}\label{eq:vKDF}
-f_{\vn+}(\vJ)&=\xi f_0'(H){(\vn\cdot\vOmega)^2\over\omega^2-(\vn\cdot\vOmega)^2}\sum_\alpha
A_\alpha\Phip_\vn(\vJ)\cr
&+g_\vn(\vJ)\delta(\omega^2-(\vn\cdot\vOmega)^2),
\end{align}
where $g_\vn$ is an arbitrary function.
Adding this term makes it possible for $f_{\vn+}$ to take whatever value $g_\vn$
specifies on the resonant surfaces. (The values taken by $g_\vn$ off resonant
surfaces are immaterial.)
Multiplying equation \eqrf{eq:vKDF} by
$\int\d^6\vw\,\e^{\i\vn\cdot\vtheta}\Phipps(\vx)$ and summing over
$\vn$, we get
\begin{align}\label{eq:givesA}
\cE &A_{\alpha'}=\int\d^3\vJ\d^3\vtheta\,
\sum_\vn\e^{\i\vn\cdot\vtheta}\Phipps(\vx)\cr
&\times\bigg\{
\xi f_0'(H){(\vn\cdot\vOmega)^2\sum_\alpha
A_\alpha\Phip_\vn\over\omega^2-(\vn\cdot\vOmega)^2}
+g_\vn(\vJ)\delta(\omega^2-(\vn\cdot\vOmega)^2)\bigg\}\cr
&=(2\pi)^3\!\cP\!\!\int\d^3\vJ\!\sum_\vn\bigg\{\xi f_0'\,
{(\vn\cdot\vOmega)^2\over\omega^2-(\vn\cdot\vOmega)^2}\sum_\alpha
A_\alpha\Phipps_\vn\Phip_\vn\cr
& +\Phipps_\vn
g_\vn(\vJ)\delta(\omega^2-(\vn\cdot\vOmega)^2)\bigg\},
\end{align}
 where the integral over $\vJ$ is a principal value in the sense that actions
at which $\vn\cdot\vOmega=\pm\omega$ are to be excluded and the large values
of the integrand as such points are approached largely cancel during
integration.  Equation \eqrf{eq:givesA} has the form
\[\label{eq:MAeqB}
\sum_\alpha \cM_{\alpha'\alpha}A_\alpha=-B_{\alpha'},
\]
where\footnote{ Our matrix $\cM$ is analogous to the matrix $\vepsilon$ of
\cite{Hamilton2018} rather than their $\vM=\vI-\vepsilon$. The integrand in our
$\cM$ differs from that of \cite{Hamilton2018} in that frequencies occur
squared because we are working with an operator that is second- rather than
first-order in time. The polarisation and response operators defined in
Chapter 5 of
\cite{GDII} also involve frequencies rather than their squares. Similar
operators can be derived from  our $\cM$ by decomposition of our integral by
partial fractions.}
\begin{align}\label{eq:defsMB}
&\cM_{\alpha'\alpha}(\omega)\equiv\delta_{\alpha'\alpha}\cr
&\ -{(2\pi)^3\xi\over\cE}\cP\!\int\d^3\vJ\,f_0'(H)\,\sum_\vn
{(\vn\cdot\vOmega)^2\over\omega^2-(\vn\cdot\vOmega)^2}
\Phipps_\vn\Phip_\vn\cr
&B_{\alpha'}\equiv-{(2\pi)^3\over\cE}\int\d^3\vJ\sum_\vn\Phipps_\vn(\vJ)
g_\vn(\vJ)\delta(\omega^2-(\vn\cdot\vOmega)^2).\cr
\end{align}

We will see below that in the case of a stable system, $\cM(\omega)$ has an
inverse for any real $\omega$. Consequently, for any real $\omega$ and $B_\alpha$ there is
always a unique corresponding $A_\alpha(\omega)$. Hence given $\omega$ and $\vn$, we
can determine the $A_\alpha$ for any function $g_\vn(\vJ)$ on the surface
$\vn\cdot\vOmega=\omega$. By equation \eqrf{eq:defsAp}, these coefficients
describe the spatial structure that $g$ generates:
\[
\Phi_A(\vx)=\xi\sum_\alpha A_\alpha\Phip(\vx)\quad;\quad
\rho_A(\vx)=\sum_\alpha A_\alpha\rhop(\vx).
\]
The $B_\alpha$ turn out to be the analogous expansion coefficients  of the density
\[
\rho_B\equiv\int\d^3\vv\,\sum_\vn
g_\vn(\vJ)\e^{\i\vn\cdot\vtheta}\delta(\omega^2-(\vn\cdot\vOmega)^2).
\]
Indeed,
\begin{align}
-{1\over\cE}\int\d^3\vx\,\Phips(\vx)\rho_B
&=-{1\over\cE}\sum_{\vn\vn'}\int\d^3\vJ\int\d^3\vtheta\,\Phips_{\vn'}(\vJ)
\e^{-\i\vn'\cdot\vtheta}\cr
&\qquad\times g_\vn(\vJ)\e^{\i\vn\cdot\vtheta}\delta(\omega^2-(\vn\cdot\vOmega)^2)\cr
&=B_\alpha.
\end{align}
We shall call modes with real frequencies and non-zero $g_\vn$ {\it van
Kampen} modes.

At $\omega^2<0$, $B_\alpha=0$ because the
argument of the $\delta$-function in the definition of $B_\alpha$ cannot vanish.
So given $\omega^2<0$ an associated DF can be found only if $|\cM|=0$. Thus
there may be isolated pure imaginary frequencies $\pm\i\omega_0$ at which a
perturbation can grow/decay exponentially. We call such modes {\it classical
modes}. They are distinct from damped Landau `modes', which have frequencies
below the real axis and not on the imaginary axis. 

Note that to prove the system's stability it suffices to show that $|\cM|$
has no zeroes on the imaginary axis. In any case, the normal-mode frequencies
are confined to a continuum of real values (van Kampen modes) with the
possible addition of discrete pure imaginary values (classical modes).

\subsection{Relation to Landau modes}

In a conventional normal-mode analysis we derive an equation $M\va=0$ which
is homogeneous in the disturbance's amplitude $\va$ with the consequence that
non-trivial solutions exist only when the determinant of the matrix $M$
vanishes. The dispersion relation is the condition for $|M(\omega)|$ to
vanish. Our recognition that there can be non-trivial distributions of stars
on resonant tori gives rise to non-vanishing $B_\alpha$, and thus causes
$A_\alpha$ to satisfy an inhomogeneous equation analogous to $M\va=\vb$ that
can be satisfied whenever $B_\alpha\ne0$, that is, at any frequency
$\omega=\vn\cdot\vOmega(\vJ)$ for some $\vn$ and $\vJ$.
Hence there is no dispersion relation associated with the true modes of a
stellar system. In general there will be complex
frequencies $\omega_0$ at which $|\cM(\omega_0)|=0$, but unless $\omega_0$
lies on an axis of the complex plane, it cannot be the frequency of a true
mode because all true modes have real $\omega^2$.

Landau modes occur at the frequencies $\omega_0$ at which a matrix
$M(\omega)$ has vanishing determinant. The Landau matrix $M$ is closely
related to $\cM$, and if $|M(\omega_0)=0$ then $|\cM(\pm\omega_0)|=0$ also. Thus
every Landau mode is associated with the possibility of solving equation
\eqrf{eq:MAeqB} with $B_\alpha=0$. Yet when $\omega_0$ lies on neither axis of the
complex plane,  such a solution should not be included in the set of true
modes for two reasons:

\begin{itemize}
\item[i)] The solution cannot be an eigenfunction of $K$ because $\omega_0^2$ is not
real, so it falls outside the complete set formed by the true modes. (It
follows that it can be written as a sum of the true modes.)

\item[ii)] The solution associated with one of $\pm\omega_0$ will grow
exponentially, so every system would be unstable if these solutions were included in the
complete set of true modes.
\end{itemize}

If a system is stable, the frequencies of its Landau modes all lie below the
real axis. In Section \ref{sec:DF} we used this fact to argue that given any
real $\omega$, equation \eqrf{eq:MAeqB} has a unique solution for $A_\alpha$
given $B_\alpha$. Time-reversal symmetry is responsible for $|\cM|$ vanishing
above the real axis whenever it vanishes below the real axis and the failure
of the determinant of the Landau matrix $M$ to behave in the same way is a
consequence of the the violation of time-reversal symmetry inherent in the
initial-value problem that leads to $M$.

\subsection{Modes and dressing}\label{sec:dress}

To understand the physical reality that underlies this mathematics, consider
that in the absence of self gravity ($\xi =0$), a non-trivial distribution of
stars with respect to $\vtheta$ on resonant surfaces will generate
oscillations in the density at frequency $\omega$ that will persist for ever.
The $\delta$-function component of the DF \eqrf{eq:vKDF} of a van Kampen mode
represents this phenomenon, and the $B_\alpha$ quantify the spatial form of
this driving structure.  When $\xi >0$, these oscillations affect the
dynamics of all stars, including non-resonant stars. The regular part of the
mode's DF \eqrf{eq:vKDF} describes these sympathetic oscillations of
non-resonant stars. The $A_\alpha$ quantify the spatial structure of this
``dressed'' response to the driver $g$. 

The functions $g_\vn$ are arbitrary, so we may consider the case in which
$g_\vn$ vanishes for all vectors but one, $\vN$. Consider now the effect of
multiplying equation \eqrf{eq:vKDF} by
$\int\d^6\vw\,\e^{\i\vn\cdot\vtheta}\Phipps(\vx)$ as in the derivation of
equation \eqrf{eq:givesA} but now {\it not} summing over $\vn$. Then we have 
\begin{align}\label{eq:no_sum}
C_{\alpha'\vn}=(2\pi)^3\!\cP\!\!\int\d^3\vJ\,\!\Phipps_\vn&\bigg\{\xi f_0'\,
{(\vn\cdot\vOmega)^2\over\omega^2-(\vn\cdot\vOmega)^2}\sum_\alpha
A_\alpha\Phip_\vn\cr
& +
g_\vn(\vJ)\delta(\omega^2-(\vn\cdot\vOmega)^2)\bigg\},
\end{align}
where
\[
C_{\alpha'\vn}\equiv-
\int\d^6\vw\,\Phips(\vx)\e^{\i\vn\cdot\vtheta}f_{\vn+},
\]
so
 \[\label{eq:AfromC}
A_\alpha={1 \over\cE}\sum_\vn C_{\alpha\vn}
\]
When $\vn\ne\vN$, the right side of equation \eqrf{eq:no_sum}
has only the term proportional to $\xi$, so would vanish with $\xi $. That is, without self-gravity,
$C_{\alpha\vn}\ne0$ only for $\vn=\vN$; in this case the van Kampen modes
could be labelled by $\vN$. In the presence of self-gravity, we have no
reason to expect $C_{\alpha\vn}$ to vanish for $\vn\ne\vN$ because equation
\eqrf{eq:AfromC} includes a contribution to  $A_\alpha$ from
$C_{\alpha\vN}\ne0$. Self-gravity has this impact because it
stops $K$ commuting with the operators $[.,J_r]$ and $[.,L]$ as discussed above. From the
fact that $K$ does commute with the operator $[.,L_z]$ it
follows that even when $\xi >0$, $A_{\alpha\vn}=0$ unless
$n_3=N_3$.

When a Landau mode is weakly damped, $|\cM|=0$ at a frequency $\omega_0$ that
lies just below the real axis, and in consequence $|\cM|$ is small on the
real axis just above this zero. In view of equation \eqrf{eq:MAeqB},
$|\vA|/|\vB|$ will be large in these circumstances. That is, van Kampen modes
with frequencies close to $\real(\omega_0)$ are ``heavily dressed''.  This is
the true significance of Landau modes.  Put differently, while van Kampen
modes exist at any $\omega$, they have a bigger footprint in action space at
frequencies that lie close to those of Landau modes.

Whereas a gas ball has at most a finite number of normal modes  at a
countable number of frequencies, a cluster has an infinite number of normal
modes at every frequency. This difference is a consequence of the likely
completeness of normal modes in each system (Section~\ref{sec:complete}) and
the fact that much more information is required to specify the DF of a
cluster than the state of a gas ball:  the disturbance has adiabatically
deformed the latter from its equilibrium, so the velocity distribution
remains Maxwellian and one only has to specify a mean and dispersion at each
location $\vx$. In a cluster we need to specify the DF in six-dimensional
phase space.

Whereas the Hermiticity of $K$ ensures that van Kampen modes for different
$\omega$ are orthogonal, it falls to us to select from all modes for any
given $\omega$ a complete set of mutually orthogonal modes. That is, 
to select a set of functions $g(\vJ)$ that generate  modes that are
orthogonal in the sense $\la g_i\ket{g_j}=\delta_{ij}$. Since $K$ commutes
with $[.,L_z]$, we can require modes to be eigenfunctions of
this Hermitian operator, and identification of a complete set of modes is reduced to finding an
orthogonal set of vectors whose components are indexed by $n_1$ and $n_2$.
Rather than solving this problem, in
Paper III we show how an arbitrary state of a stellar system can be
decomposed into its constituent modes. That is, to express a given
perturbation at time zero, $F(\vw,0)$, in the form
\[ \label{eq:gen_sum}
F(\vw,0)=\int\d\omega\,f(\vw,\omega),
\]
where $f(\vw,\omega)$ is a van Kampen mode with frequency $\omega$. This
done, the state of the perturbation at any other time can be obtained as
\[
F(\vw,t)=\int\d\omega\,f(\vw,\omega)\,\e^{-\i\omega t}.
\] 
These integrals over $\omega$, eliminate the principal-value and Dirac
$\delta$-function symbols in equation \eqrf{eq:vKDF} for $f$.

\subsection{Energy  of a  mode}\label{sec:innerProd}

Equation \eqrf{eq:EvK} says that  the energy of a mode  is just the
norm of the odd part of its
DF, and equation \eqrf{eq:vKDF} gives the even part of a mode's DF from which the
odd part follows trivially. So the natural next step is to substitute from
the second of these equations  into the first and express the mode's energy
in terms of its parameters  $g_\vn$ and its potential $\Phi[f]$.  This
exercise proves long, and is made  intricate by the singular  denominator
$\omega^2-(\vn\cdot\vOmega)^2$ in equation \eqrf{eq:vKDF} for the DF.

In Appendix \ref{app:ff}
we compute $\la f\ket{\widetilde f}$ for modes with frequencies $\omega$ and
$\widetilde\omega$.
The result is\footnote{In equation \eqrf{eq:dotvKf} $n_3$ has the same, fixed
value for both $f$ and $\widetilde f$.}
\begin{align} \label{eq:dotvKf}
\la f_-&\ket{\widetilde f_-}=
-(2\pi)^3\,\delta(\omega^2-\widetilde\omega^2)\int\d^3\vJ
\sum_{n_1,n_2}\cr
&\bigg(\pi^2\omega^4f_0'
\Phi^*_\vn[f]\Phi_\vn[\widetilde f]+{g_\vn^*\widetilde g_\vn\over f_0'}\bigg)
\,\delta((\vn\cdot\vOmega)^2-\omega^2).
\end{align}
A pleasing feature of this expression is that it has no reference to
potential/density basis functions. The factor
$\delta(\omega^2-\widetilde\omega^2)$ on the rhs reflects the orthogonality of
modes of different frequencies that follows from the Hermiticity of
$K$.\footnote{The appearance of $\omega^2$ rather than $\omega$ reflects
time-reversal symmetry. From a practical perspective, it ensures that real
time dependence ($\cos\omega t$) is possible.} A remarkable feature of
equations \eqrf{eq:dotvKf} is that its action-space integral is confined to
the resonant tori, even though the modes very much involve non-resonant
stars.

When we set $\widetilde f=f$, the inner
product is divergent for finite $g$ because finite $g$ generates an $f_+$
that diverges as the resonant surface in action space is approached, and the
divergences on opposite sides of the surface do not cancel because energy
density is proportional to $f^2$. This result signals that we can have only
an infinitesimal number of stars on any resonant surface. When use equation
\eqrf{eq:dotvKf} to compute the
energy of a physical disturbance \eqrf{eq:gen_sum}, we find
\begin{align}\label{eq:EvKn}
E[F]&=\bra{F_-}F_-\ra=\Big<{\int\d\omega\,
f_-(\omega)}\Big|\int\d\omega'\,f_-(\omega')\Big\ra\cr
&=\int\d\omega\,\d\omega'\bra{f_-(\omega)}f_-(\omega')\ra\cr
&=-(2\pi)^3\sum_{\vn}\int{\d\omega\over2\omega}\int\d^3\vJ
\bigg(\pi^2\omega^4f_0'
|\Phi_\vn[f]|^2+{|g_\vn|^2\over f_0'}\bigg)\cr
&\hskip3cm\times\delta((\vn\cdot\vOmega)^2-\omega^2)\cr
=-&{(2\pi)^3\over4}\sum_{\vn}\!\int\!\d^3\vJ
\bigg(\pi^2(\vn\cdot\vOmega)^2f_0'
|\Phi_\vn[f]|^2\!
+{|g_\vn|^2\over(\vn\cdot\vOmega)^2 f_0'}\bigg).\cr
\end{align}
There is a striking similarity between the first term in the expression
\eqrf{eq:EvKn} for $E$ and the expression $\rho_E=\fracj12\omega^2\epsilon_0A^2$
for the energy density contributed by an electromagnetic wave with
vector-potential amplitude $A$. In the electromagnetic case one factor of
$\omega$ arises from the quantisation condition $E=\hbar\omega$ and the other
arises from the canonical momentum $\omega\vA$ of the field $\vA$. The second
term in equation \eqrf{eq:EvKn} has a different structure, however, and this term is
arguably more important than the first because $\Phi[f]$ is driven by
$g$.

A natural question to ask is why the coefficient of $\big|\Phi_\vn[f]\big|^2$
in equation \eqrf{eq:EvKn} is positive, given that gravitational potential energy is
inherently negative. The explanation must be that this term encapsulates all
the energy, kinetic as well as potential, that's tied up in the disturbance
in non-resonant stars that is excited by  the resonant stars. In the absence
of self-gravity, the non-resonant stars are not disturbed, so this
contribution to the energy vanishes with $\Phi[f]$.

The second term in the integrand of equation \eqrf{eq:EvKn} is ultimately
limited  by
Poisson noise and can be considered a given, while the first term depends on
the system's dynamics.  The ratio of the two terms is proportional to
$(\vn\cdot\vOmega)^4f_0^{\prime2}$, so the relative contributions to $E$ from
the resonant driver $g$ and the non-resonant response $\Phi$ are sensitive to
this factor. In principle $|\vn\cdot\vOmega|$ can be made as small as we
please at given $\vJ$, but only by going to large $|\vn|$, and at large
$|\vn|$ the projection of $g_\vn(\vJ)$ into real space (measured by
$B_\alpha$) will be small and thus the response (measured by $A_\alpha$ and
$\Phi[f]$) will
be small.  Hence the first term in the integrand of equation \eqrf{eq:EvKn} will be
significant only for small $|\vn|$. The fundamental dipole mode most
obviously satisfies this criterion.  

This consideration draws attention to short vectors $\vn$ that make
$|\vn\cdot\vOmega|$ small (if it is small on any torus, it will be small on
many tori) because for these vectors the noise component $g_\vn$ generates
a response at the least cost in energy, so the response is likely to be
large. We saw above, moreover, that the response is enhanced when
$\vn\cdot\vOmega$ is close to the real part of the frequency of a weakly
damped Landau mode, because then $|\cM|$ is small.  The fundamental
dipole mode has been shown to be weakly damped in typical models
\citep{Weinberg1994,
Saha1991, Hamilton2018}.

\subsection{Initialisation of N-body models}\label{sec:Nbody}

Suppose we set up a cluster by randomly sampling an analytic DF. When the
selection is complete, the actual DF will differ from the analytic DF by
virtue of Poisson noise, so the $g_\vn$ will be non-zero and the cluster's
van Kampen modes will be excited.  The coefficients $B_\alpha$ that quantify
the noisiness of the density distribution are unambiguously fixed by the
Monte-Carlo selection, but the potential 
that are generated
from them $\xi\sum_{\alpha\beta}\Phi^{(\alpha)}\cM^{-1}(\xi,\omega)B_\beta$
will vary with the degree of
self-gravity $\xi $. Hence the DFs of the modes that sum to the sampled
phase-space distribution will depend on $\xi $, but their sum must produce
the DF sampled regardless of $\xi $. When $\xi =1$, the modes are heavily
dressed and yet produce the same small (Poisson) fluctuations in density as
in the case $\xi =0$ of vanishing self-gravity because the contributions of
different modes cancel to a considerable extent.  This cancellation is
particularly pronounced in the case of low-order modes (small $|\vn|$), and
it occurs because when $\xi =1$ the phases of modes are correlated, whereas
when $\xi =0$ they are probably uncorrelated.

Once we start moving the stars with $\xi $ set to unity, the phase
differences between modes with different frequencies will tend towards
uniform distribution in $(0,2\pi)$ and cancellations between perturbations to
the density will diminish. Consequently, the heavily dressed individual
modes will become manifest and the system will  become less spherical
as Lau \& Binney (2019) found empirically. By contrast, when
stars are moved in the analytic potential, the initially uniformly
distributed phases of the modes obtained with $\xi =0$ remain uniformly distributed and
no significant change in the density fluctuations will be observed. 

The larger the value of $\xi $, the larger will be the values of $A_\alpha$
that are generated by the given $B_\alpha$, so the greater will be the
departures from spherical symmetry once the phases of modes have
decorrelated.

To obtain a self-consistent realisation of a self-gravitating system one
needs to excite the modes for $\xi =1$ with random phases, and it is not clear
how this can be done without computing the system's modes.

\section{Completeness of modes}\label{sec:complete}

We have defined the true modes of a stellar system to be the
eigenfunctions of Antonov's Hermitian operator $K$. In quantum mechanics it
is conventional to assume that the eigenfunctions of any Hermitian
operator form a complete set although proof of completeness requires the
operator to be bounded \citep[e.g.][\S11.5]{Dieudonne}, which some operators of physical interest
are not. Similarly, much of condensed-matter physics relies on Bloch's
theorem that there is a complete set of stationary states for an electron in
a crystal that have wavefunctions of the form
$\psi(\vx)=\e^{\i\vk\cdot\vx}u(\vx)$ with $u(\vx+\va)=u(\vx)$ for any
lattice vector $\va$. The standard derivation of Bloch's theorem
\citep[e.g.][\S14.4]{ElliottDawber} starts from the observation that if $\psi(\vx)$
is a stationary state, then so is $\psi(\vx+\va)$. Hence the stationary states of a given energy provide a
representation of the Abelian translation group. Such groups only have
one-dimensional irreducible representations, so the action of the group can
be reduced to multiplication by $\e^{\i\vk\cdot\va}$. That is, any functions
providing a representation of the translation group can be reduced to
functions satisfying
$\psi(\vx+\va)=\e^{\i\vk\cdot\va}\psi(\vx)$, a relation that is clearly satisfied by
$\psi(\vx)=\e^{\i\vk\cdot\vx}u(\vx)$. 

Analogously, we might argue that the time-translation
invariance of $\p^2_t+K$ implies that if $f(\vw,t)$ satisfies
$(\p^2_t+K)f=0$, then so does $f(\vw,t+\tau)$ and it follows that any set of
solutions can be reduced to ones that satisfy
$f(\vw,t+\tau)=\e^{-\i\omega\tau}f(\vw,t)$. Then setting $t$ to zero we
infer that solutions of the form $f(\vw,\tau)=\e^{-\i\omega\tau}f(\vw)$ are
complete.

The above arguments for the completeness of Bloch waves and eigenfunctions of
$A$
are open to the objection that the theorem regarding the decomposition of
representations into irreducible representations requires the group to be
compact, which translation groups are not.\footnote{The reduction theorem
applies only to unitary representations, which associates every group member
$g$ with a unitary operator $T_g$ on a vector space. If a group is compact,
Maschke's operator $S^2=\sum_gT^\dagger_gT_g$ can be used to establish that
any representation is isomorphic to a unitary representation. In the
non-compact case the sum over $g$ is ill-defined.} Hence, the completeness of
van Kampen modes cannot be rigorously established by the group-theoretic
argument, although similar arguments are widely accepted in physics.

\cite{Case1959} established the completeness of the van Kampen modes of an
electrostatic plasma by direct demonstration that any DF
$f(\vw)$ can be written as a sum of van Kampen modes. The corresponding
exercise for stellar systems will be presented in Paper III of this series.

\section{Discussion}\label{sec:discuss}

\subsection{Particle dressing}\label{sec:dressing}

The concept of particle dressing has been central to high-energy physics for
over half a century, but has been slow to catch on in stellar dynamics. In
galactic dynamics it can be traced at least as far back as \cite{JT1966}, who
showed that a gas cloud in a galactic disc would attract an entourage of
passing stars $\sim$ten times more massive than itself. \cite{Toomre1991}
showed that individual disc stars also enhance their masses tenfold by
attracting an (ever-changing) entourage of other stars.
\cite{SellwoodCarlberg2014} showed that large-amplitude spiral structure
emerges from Poisson noise through successive spiral instabilities, but this
important process was only firmly connected to particle dressing by
\cite{FouvryPMC2015}. \cite{Hamilton2021} made the connection between the BL
equation and particle dressing beautifully clear via Rostoker's principle:
that it is permissible to compute the effects of discreteness as from the
interaction of uncorrelated but dressed particles \citep{Rostoker1964}. Here
we have interpreted a van Kampen mode as the result of dressing not one star
but an ensemble of stars on a group of resonant tori. 

Given that the CBE is the first equation in the BBGKY hierarchy of equations
with the two-particle correlation function set to zero, the importance of
dressing for the structure of van Kampen modes may seem paradoxical. The CBE is a
mean-field approximation akin to the Weiss theory of magnetism, and embraces
correlations that are induced by perturbing fields. Hence it embraces the
dressing of resonant tori involved in van Kampen modes. Moreover, when a
simulation is started, its DF is inevitably perturbed from the underlying
analytic DF and thus its van Kampen modes are excited.

\subsection{van Kampen vs Landau modes}\label{sec:vK_Landau}

Normal modes are perhaps the most important single tool in theoretical
physics -- quantum field theory has even taught us to see particles as
excitations of normal modes of the vacuum. Modal analyses have played a
significant role in stellar dynamics since the seminal work of
\cite{Kalnajs1965}, Toomre \citep{To64,Toomre1981} and later the prescient
work of Weinberg \citep{Weinberg1993,Weinberg1994,Weinberg1998,Weinberg2001},
but in all these studies the modes considered were those of Landau. These
`modes' lack key properties of true modes: (i) completeness in the sense that
any initial condition can be expressed as a linear combination of modes, and
(ii) additivity of energies. These two properties are essential for the use
of modes in physics and engineering outside stellar dynamics. The contents
pages of the two volumes of \cite{FridmanPolyachenkoI} explain the focus on
Landau modes: the community wanted to establish which equilibrium models are
stable, rather than to investigate, as we do, the excursions that stable
systems make around equilibrium. 

Doubt is sometimes cast on the physical standing of van Kampen modes because
their DFs contain a $\delta$-function. Actually this feature is a natural
consequence of their forming a continuum. Testable predictions
of van Kampen modes will always emerge after integration over $\omega$, just
as in the familiar  quantum-mechanical treatment of radiative transitions
sensible results emerge only after integration over the frequency of the
electromagnetic field,\footnote{When deriving Fermi's golden rule, one
analogously integrates over the energies of final states.} and the
$\delta$-functions will disappear in the process. Landau
modes are superpositions of van Kampen modes, and they decay as their
constituent van Kampen modes drift apart in phase \citep{Case1959}. If you
run a decayed Landau mode back in time, the phases move back into alignment
for a finite time before drifting apart, so the disturbance grows for only a
limited time.

\cite{HamiltonHeinemann2020} take a fresh approach to relaxation in stellar
systems that involves Landau modes in an essential way.
\cite{BinneyLacey1988} showed that the  diffusion tensor of the action-space
Fokker-Planck equation follows immediately from the temporal power spectrum of the
gravitational potential. \cite{HamiltonHeinemann2020} argue that both dressed
two-particle interactions and normal modes of the entire system contribute to
the power spectrum. They assume that the modes in question are Landau modes,
which they imagine to reach an equilibrium level of excitation through their
native damping being offset by constant excitation by Poisson noise.  This
picture involves a transfer of energy from Landau modes to the underlying
heat bath, and then back to the Landau modes via Poisson noise. The mechanism
by which Poisson noise draws energy from the heat bath is unclear. 

In a simpler picture each van Kampen mode has a fixed amplitude and energy
and a phase that advances at its own steady rate. Modes with frequencies that
lie close to the real parts of a weakly damped Landau mode have large
amplitudes because they are heavily dressed.  From time to time their phases
yield constructive interference and the Landau mode appears to be highly
excited. The excitation decays as shifts in relative phase spoil the
constructive interference. At a later time the phases again align favourably,
and the process repeats.

\subsection{Thermodynamics of star clusters}\label{sec:thermo}

When a stellar system is born, its van Kampen modes are assigned particular
amplitudes and phases. At birth the phases may be highly correlated, but they
will decorrelate on a dynamical timescale. This decorrelation may manifest
itself through the emergence of system-scale fluctuations in the density,
wandering of the point of highest central density, etc
\citep{LauBinney2019,Heggie2020}.  On a longer timescale non-linear terms in
the CBE will mediate exchanges of energy between modes. We know that the
amplitudes of modes are invariant at linear order, and we expect them to
evolve at quadratic order in the perturbations. The two-body timescale is
precisely the timescale associated with terms of quadratic oder
\citep{Chavanis2012}, so van Kampen modes are expected to exchange energy on
the two-body timescale. 

There is no reason to believe that when a cluster is first realised the
amplitudes of its van Kampen modes conform to the Gibbs distribution.  We
expect exchanges of energy between modes to drive the distribution towards
the Gibbs distribution and equipartition of energy between
modes, which is to say that the actual DF is $F=\int\d\omega\,f(\omega)$ with
$f(\omega)$ of van-Kampen form \eqrf{eq:vKDF} and (cf. eqn~\ref{eq:EvKn})
\[
E[f]=\bra{f_-}f_-\ra=\hbox{constant}.
\]
Formally, modes exchange energy and equipartition
can be approached on the same (two-body) timescale on which core collapse
and evaporation change the mean-field model, but since core collapse
occurs after $\sim300$ central two-body times \citep[][\S7.5.3]{GDII}, it is
plausible that a good approximation to the Gibbs distribution will be
achieved before core collapse occurs. Hence a seductive programme of work is
to assume equipartition of energy and random phases between the modes of a
particular mean-field model and to compare the observables predicted thus
with observational data and N-body simulations.  

In this regard it is instructive to compare the applicability of
thermodynamics to star clusters and to classical systems that also have
access to states of very high entropy, for example a mixture of two parts
hydrogen and one part oxygen, or a diamond, both of which have higher entropy
states (water vapour and graphite) that can only be reached by climbing over
a significant energy barrier. On account of this barrier, a hydrogen/oxygen
mixture and a diamond will extensively explore the configurations accessible
with thermal energy regardless of the existence of states of much higher
entropy. A stellar system is not denied access to states of high energy by an
energy barrier but by a glacial rate of energy transport.

Nevertheless, computing the thermodynamics of a cluster is feasible because
the positivity of model energy ensures that the cluster's constant-energy
surfaces have finite volume $\Omega$, and it would be very interesting to
examine its predictions. 

In such a theory the DF would itself be a random variable in addition to the
coordinates of stars, which are the random variables whose probability
distribution the DF specifies. Testable predictions would emerge from the
theory as double expectations: first
$\ex{\cO}_{f}=\int\d^6\vw\,f(\vw)\cO(\vw)$ and then an average of these
averages weighted by the probability of each DF $f$. Hence a prerequisite of
the theory is the ability to assign probabilities to DFs in a rational way.
In particular, the probability assigned to a group of DFs must remain
unchanged as a cluster evolves under the CBE. In standard statistical
mechanics the
analogous requirement, that a probability density on phase space be invariant
under Hamiltonian evolution, is satisfied by making a priori probability
proportional to the measure of phase-space volume $\d^N\vq\,\d^N\vp$ that
canonical coordinates $(\vq,\vp)$ deliver.  In a companion paper
\citep{LauBinney_probDF} we extend this
idea to the space of distribution functions by defining canonical coordinates
for this space. It turns out that the energy of a van Kampen mode then takes the form
of a sum of Hamiltonians of simple-harmonic oscillators.

\subsection{Prior work}\label{sec:prior}

To our knowledge the van Kampen modes of a stellar system have previously
been considered only by \cite{Vandervoort2003}, who followed
\cite{vanKampen1955} in deriving the modes directly from the CBE.
\cite{Antonov1961} obtained the second-order, Hermitian differential operator
$K$ by splitting the DF into parts even and odd in $\vv$, but he was focused
on proving his stability principle and didn't show that the eigenfunctions of
$K$ are the van Kampen modes. He didn't take advantage of angle-action
variables or compute excitation energies. \cite{Polyachenko2021} discussed
the relation of Landau and van Kampen modes in the context of the periodic
cube, though principally in the unphysical case that the cube's mass exceeds
the Jeans mass so the system is unstable. Their work makes very clear that
Landau `modes' lack essential properties of modes, and also illustrates what
a treacherous arena the complex plane is: physically ill-motivated changes in
the contour of integration over velocity give rise to solutions with
radically different properties. In particular they show that Landau's choice
of contour breaks the time-reversal symmetry of the underlying problem. 

\cite{Polyachenko2021} introduced the nomenclature `true mode' for a member
of the complete set of modes and like us restricted the term `van Kampen
mode' to true modes with real frequencies. The principal point made by
\cite{Polyachenko2021} is that the unstable Jeans mode at $\omega=\i y$ is
accompanied by a decaying mode at $\omega=-\i y$. This is a trivial consequence of
time-reversibility but \cite{Polyachenko2021} show that some effort is
required to explain why Landau's analysis misses this mode.

\section{Conclusions}\label{sec:conclude}

As the completeness and precision of astronomical data grow, the oscillations
of stellar systems around equilibrium configurations will increase in
observational significance. The natural way to produce theoretical
predictions of these phenomena is to adapt the techniques of statistical
mechanics to stellar dynamics. This paper takes a step in this direction for
ergodic stellar systems by focusing attention on the
van Kampen modes of stellar systems, which have hitherto been eclipsed by
Landau modes.

We showed for the first time that van Kampen modes of an ergodic system are
the eigenfunctions with positive eigenvalues of Antonov's second-order
Hermitian operator on phase space. In consequence, the true modes of an
ergodic stellar
system are either purely sinusoidal or exponentially growing/decaying; there are no
over-stable modes or modes comprising decaying oscillations. The frequencies
of oscillating modes form a continuum, and the DFs of these modes contain
$\delta$-functions which disappear when testable predictions are extracted by
integrating over frequencies. Any exponentially growing/decaying modes are
isolated in frequency space and their DFs do not contain $\delta$-functions. The
energy of an oscillating mode is just the norm of the odd part of its DF.
From this it follows that these modes have positive energy. The energy of a
growing/decaying mode is identically zero.

We interpreted van Kampen modes as dressed sets of resonant tori. How heavily
they are dressed increases with the extent to which the system is
self-gravitating, and with proximity in frequency space to a zero of the
response matrix $\cM(\omega)$ -- the zeroes of this matrix come in pairs with
each pair that is not on the imaginary axis  associated
with a Landau `mode'.  Landau modes are not members of the complete set of
true modes and hence are linear combinations of true modes.

A star cluster has many more true modes than the equivalent
gas ball because much more information is required to specify a DF than to
specify the density and pressure in a ball of gas. For this reason one
hesitates to enumerate a cluster's van Kampen modes, except possibly in the
much simplified case of vanishing self-gravity. Paper III shows that we can
avoid this enumeration by showing how to decompose any initial state $F(\vw)$
of the system into a linear combination $F=\int\d\omega\,f(\omega)$ of van
Kampen DFs $f(\vw,\omega)$.  This decomposition automatically identifies the
particular mode at frequency $\omega$  that is required
to synthesise the given DF.

 When a model cluster is realised, its van Kampen modes acquire non-zero
amplitudes by virtue of Poisson noise.  The phases of modes evolve on a
dynamical timescale while their amplitudes probably evolve on the two-body
timescale.  Consequently, there is an early phase of relaxation in the
evolution of a simulated cluster in which system-scale distortions emerge.
Consideration of the way modes depend of the degree of self-gravity explains
why system-scale  distortions are less prominent in simulations that are less
self-gravitating.

The positivity of the energies of van Kampen modes opens the door to the
application of standard statistical physics to stellar systems: while in the
long term systems will drift through core collapse and evaporation to states
of ever higher entropy, in the medium term disturbed systems may relax to
distributions of energy among van Kampen modes that maximise entropy.

\section*{Acknowledgements}

We thank an anonymous referee for an exceptionally careful reading and
numerous constructive suggestions, and we thank J.\ Magorrian and B.\ Kocsis
for helpful comments on early drafts. Our discussions of the completeness of
van Kampen modes benefited from discussions with U.\ Tillmann. Jun Yan Lau
gratefully acknowledges support from University College London's Overseas and
Graduate Research Scholarships.  James Binney is supported by the UK Science
and Technology Facilities Council under grant number ST/N000919/1 and by the
Leverhulme Trust through an Emeritus Fellowship. 

\section*{Data Availability}

No new data was generated or analysed in support of this research.

\bibliographystyle{mn2e} 
\bibliography{new_refs.bib}

\appendix

\section{Matrix elements of Antonov's operator}\label{app:fKf}

Here we compute a general matrix element $\bra{\widetilde f_-}K\ket{f_-}$ of
Antonov's operator, with $f$ and  $\widetilde f$ any two DFs. Equation 
\eqrf{eq:fddwithj} yields
\begin{align}\label{eq:gKf}
\bra{\widetilde f_-}K&\ket{f_-}=\int{\d^6\vw\over
f'_0}\,\widetilde f_-^*\left(\vOmega\cdot{\p\over\p\vtheta}\right)^2f_-\cr
&+\i{\xi\over\cE}\int\d^6\vw\,\widetilde f_-^*\vOmega\cdot{\p\over\p\vtheta}\sum_\alpha\Phip(\vx)j_\alpha[f_1].
\end{align}
Since we can write $\d^6\vw=\d^3\vJ\d^3\vtheta$, we can shift the derivatives
wrt $\vtheta$ around by partial integration and obtain
\begin{align}\label{eq:startShift}
\bra{\widetilde f_-}K\ket{f_-}&=-\int{\d^6\vw\over
f'_0}\,\left(\vOmega\cdot{\p \widetilde f_-^*\over\p\vtheta}\right)\left(\vOmega\cdot{\p
f_-\over\p\vtheta}\right)\cr
&\quad-\i\int\d^6\vw\,\sum_\alpha\Phip(\vx)\vOmega\cdot{\p
\widetilde f_-^*\over\p\vtheta}j_\alpha[f_1]\cr
&=-\int{\d^6\vw\over
f'_0}\,\left(\vOmega\cdot{\p \widetilde f_-^*\over\p\vtheta}\right)\left(\vOmega\cdot{\p
f_-\over\p\vtheta}\right)\cr
&\quad -\i{\cE\over\xi}\sum_\alpha j_\alpha^*[\widetilde f_1]j_\alpha[f_1].
\end{align}
 The symmetry of the rhs wrt $f,\widetilde f$ implies that $K$ is Hermitian.
Since $K$ is Hermitian, all its eigenvalues $\omega^2$ are real and $K$'s eigenfunctions are
either sinusoidal or show pure exponential growth/decay.

When we set
$\widetilde f=f$, we obtain
\[\label{eq:fKf}
\bra{f_-}K\ket{f_-}=\int{\d^6\vw\over
|f'_0|}\,\left|\vOmega\cdot{\p
f_-\over\p\vtheta}\right|^2 
-{\cE\over\xi}\sum_\alpha \big|j_\alpha[f_1]\big|^2.
\]
It is interesting to express the rhs of equation \eqrf{eq:fKf} in terms of $f_+$ 
using equations \eqrf{eq:splitfdot},
\eqrf{eq:PoissonHzero} and \eqrf{eq:jA}. The result is
\[\label{eq:braKket}
\bra{f_-}K\ket{f_-}=\int{\d^6\vw\over
|f'_0|}\,\left|{\p f_+\over\p t}\right|^2 
-{\cE\over\xi}\sum_\alpha \bigg|{\p A_\alpha[f_+]\over\p t}\bigg|^2.
\]
In the case that $f_-$ is an eigenfunction of $K$ with eigenvalue
$\omega^2$, we can replace time derivatives by $-\i\omega$ (with $\omega$
potentially pure imaginary), and equation
\eqrf{eq:braKket} becomes
\begin{align}
\bra{f_-}K\ket{f_-}&=\omega^2\la f_-\ket{f_-}\cr
&=|\omega^2|\left\{\int{\d^6\vw\over
|f'_0|}\,|f_+|^2
-\cEx\sum_\alpha\big|A_\alpha[f_+]\big|^2\right\}\cr
&=|\omega^2|\left\{\la f_+\ket{f_+}-\cEx\sum_\alpha\big|A_\alpha[f_+]\big|^2
\right\}.
\end{align}

\section{Inner product of van Kampen modes}\label{app:ff}

Here we compute the inner product of two van Kampen modes of the same
frequency. We have
\begin{align}
\la f_-\ket{\widetilde f_-}&=-\int{\d^6\vw\over f_0'}f_-^*\widetilde
f_-\cr
&=-(2\pi)^3\int{\d^3\vJ\over f_0'}\sum_\vn f_{\vn-}^*\widetilde f_{\vn-}.
\end{align}
 Hence $\la f_-\ket{\widetilde f_-}=0$ unless there is some vector $\vn$ at which
both $f_{\vn-}$ and $\widetilde f_{\vn-}$ are non-zero. This fact confirms that
in the absence of self-gravity, when $f_{\vn-}$ vanishes if $g_{\vn}$
vanishes, the sought-after basis modes for a given frequency are indexed by
$\vn$ in the sense that the Fourier expansions of their DFs contain only
$\pm\vn$. When $\xi >0$,
$f_{\vn'-}$ is expected to be non-zero when $g_{\vn'}=0$ providing
$g_\vn\ne0$ for a vector $\vn$ such that $n_3=n'_3$.

Equation \eqrf{eq:vKDF} gives $f_+$ for a van Kampen mode, and we have seen
that in the case of a mode
$|\vn\cdot\vOmega| f_{\vn-}=\omega f_{\vn+}$, so
\begin{align}\label{eq:massive}
&\la f_-\ket{\widetilde f_-}=-(2\pi)^3\int{\d^3\vJ\over
f_0'}\sum_{n_1,n_2}{\omega\widetilde\omega\over(\vn\cdot\vOmega)^2} f_{\vn+}^*\widetilde
f_{\vn+} \cr
&\qquad=-(2\pi)^3\omega\widetilde\omega\,\cP\!\!\int{\d^3\vJ\over
f_0'}\!\!\sum_{n_1,n_2}{1\over(\vn\cdot\vOmega)^2}\cr
&\times\bigg({(\vn\cdot\vOmega)^2f_0'\over\omega^2-(\vn\cdot\vOmega)^2}
\sum_\alpha
A^*_\alpha\Phips_\vn(\vJ)
+g^*_\vn(\vJ)\delta(\omega^2-(\vn\cdot\vOmega)^2)\bigg)\cr
&\times
\bigg({(\vn\cdot\vOmega)^2f_0'\over\widetilde\omega^2-(\vn\cdot\vOmega)^2}\sum_\alpha
\widetilde A_\alpha\Phip_\vn(\vJ)
 +\widetilde g_\vn(\vJ)\delta(\widetilde\omega^2-(\vn\cdot\vOmega)^2)\bigg),\cr
\end{align}
 where quantities associated with the mode $\widetilde f$ are marked by tildes.
When we multiply out the big brackets, we get a term with two denominators of
the form $(\vn\cdot\vOmega)^2-\omega^2$. The integral over these is to be
interpreted as a principal value, that is by excluding points at which the
denominator vanishes. We use the identity \citep[e.g.][]{RamosWhite2018}
\begin{align}
\cP{1\over x-x_1}\cP{1\over x-x_2}&=\cP{1\over x_1-x_2}\bigg(\cP{1\over
x-x_1}-\cP{1\over x-x_2}\bigg)\cr
&+\pi^2\delta(x-x_1)\delta(x_1-x_2)
\end{align}
to rewrite this term as
\begin{align}
\int\d^3\vJ\cdots=&\cP\!\int{\d^3\vJ\over
f_0'}\sum_{n_1,n_2}(\vn\cdot\vOmega f_0')^2
\sum_{\alpha\beta}A_\alpha^*\widetilde A_\beta\Phips_\vn\Phi^{(\beta)}_\vn
\cr
&\times\bigg\{{1\over\omega^2-\widetilde\omega^2} \bigg(
{1\over(\vn\cdot\vOmega)^2-\omega^2}-{1\over(\vn\cdot\vOmega)^2-\widetilde\omega^2}\bigg)\cr
&\quad+\pi^2\delta((\vn\cdot\vOmega)^2-\omega^2)\delta(\omega^2-\widetilde\omega^2)\bigg\}.
\end{align}
The cross terms in the product of equation  \eqrf{eq:massive} can be written
\begin{align}
\int\d^3\vJ\cdots&=\int{\d^3\vJ\over f_0'}
{f_0'\over\omega^2-\widetilde\omega^2}\sum_{n_1,n_2}\cr
&\times\bigg(\sum_\alpha
A_\alpha^*\Phips_\vn\widetilde g_\vn\delta((\vn\cdot\vOmega)^2-\widetilde\omega^2)\cr
&-\sum_\beta\widetilde A_\beta\Phi^{(\beta)}_\vn g_\vn^*\delta((\vn\cdot\vOmega)^2-\omega^2)\bigg),
\end{align}
while the term involving $g_\vn^*\widetilde g_\vn$ can be written
\[
\int\d^3\vJ\cdots=\int{\d^3\vJ\over f'_0}\sum_{n_1,n_2}{g_\vn^*\widetilde
g_\vn\over(\vn\cdot\vOmega)^2}
\,\delta((\vn\cdot\vOmega)^2-\omega^2)\,\delta(\omega^2-\widetilde\omega^2).
\]
Adding these fragments together and reinstating the  prefactor in equation
\eqrf{eq:massive} we have
\begin{align}
&\la f_-\ket{\widetilde f_-}=-(2\pi^3){\omega\widetilde\omega}
\,\cP\!\!\int{\d^3\vJ\over f_0'}
\sum_{n_1,n_2}\bigg\{{f_0'\over\omega^2-\widetilde\omega^2}\bigg[\cr
&\times
\bigg({(\vn\cdot\vOmega)^2 f_0'\over(\vn\cdot\vOmega)^2-\omega^2}
\sum_\alpha
A_\alpha^*\Phips_\vn+g_\vn^*\delta((\vn\cdot\vOmega)^2-\omega^2)\bigg)
\sum_\beta\widetilde A_\beta\Phi^{(\beta)}_\vn\cr
&-
\bigg({(\vn\cdot\vOmega)^2f'_0\over(\vn\cdot\vOmega)^2-\widetilde\omega^2}
\sum_\beta\widetilde A_\beta\Phi^{(\beta)}_\vn
+\widetilde g\delta((\vn\cdot\vOmega)^2-\widetilde\omega^2)\bigg)
\sum_\alpha
A_\alpha^*\Phips_\vn\bigg]\cr
&+\bigg(\pi^2(\vn\cdot\vOmega f_0')^2\sum_{\alpha\beta}
A_\alpha^*\widetilde A_\beta\Phips_\vn\Phi^{(\beta)}_\vn
+{g_\vn^*\widetilde g_\vn\over(\vn\cdot\vOmega)^2}\bigg)\cr
&\qquad\times\delta((\vn\cdot\vOmega)^2-\omega^2)\,\delta(\omega^2-\widetilde\omega^2)\bigg\}\cr
&=-(2\pi)^3{\omega\widetilde\omega\over\omega^2-\widetilde\omega^2}\int\d^3\vJ\,\sum_{n_1,n_2}
\bigg(f_{\vn+}^*\sum_\beta\widetilde A_\beta\Phi^{(\beta)}_\vn\cr
&\qquad\qquad-\widetilde f_{\vn+}\sum_\alpha
A_\alpha^*\Phips_\vn\bigg)\cr
&-(2\pi)^3{\omega\widetilde\omega}\int\d^3\vJ\,\sum_{n_1,n_2}
\bigg(\pi^2(\vn\cdot\vOmega)^2 f_0'\sum_{\alpha\beta}
A_\alpha^*\widetilde A_\beta\Phips_\vn\Phi^{(\beta)}_\vn\cr
&\qquad+{g_\vn^*\widetilde g_\vn\over(\vn\cdot\vOmega)^2f_0'}\bigg)
\delta((\vn\cdot\vOmega)^2-\omega^2)\,\delta(\omega^2-\widetilde\omega^2).
\end{align}

Now
\begin{align}
A_\alpha&=\int\d^6\vw\,\Phips f_+\cr
&=\int\d^3\vJ\,\d^3\vtheta\sum_\vn\Phips_\vn
\e^{-\i\vn\cdot\vtheta}f_+=\sum_\vn\Phips_\vn f_{\vn+},
\end{align}
 so the first integral in our final expression for $\la f_-\ket{f_-}$ vanishes
because its integrand is $\sum_\alpha(\widetilde A_\alpha A_\alpha^*-A_\alpha^*\widetilde
A_\alpha)$. With some further simplifications we can write the inner product
in the form given by equation \eqrf{eq:dotvKf}.

\end{document}